\pdfoutput=1

\documentclass[11pt]{article}

\usepackage[preprint]{acl}

\usepackage{times}
\usepackage{latexsym}

\usepackage[T1]{fontenc}

\usepackage[utf8]{inputenc}

\usepackage{microtype}

\usepackage{inconsolata}

\usepackage{graphicx}
\usepackage{authblk}
\usepackage{amsmath, amssymb}
\usepackage{booktabs}
\usepackage{hyperref}
\usepackage{multirow}
\usepackage{enumitem}
\usepackage[table]{xcolor}
\usepackage{pifont}

\title{\textit{DynaGraph}: Lightweight Multi-Model Interaction Framework via Dynamic Topological Reconfiguration}

\makeatletter
\renewcommand{\@fnsymbol}[1]{%
  \ensuremath{%
    \ifcase#1
      \or *    
      \or \dagger 
    \fi
  }%
}
\makeatother

\author{
  \textbf{Yanxing Guo$^{1,}$}\thanks{means equal contribution.},
  \textbf{Zihao Zheng$^{1,*}$},
  \textbf{Fangzhou Wu$^{2}$},
  \textbf{Ling Liang$^{1,3,}$}\thanks{indicates the corresponding author.},
\\
  \textbf{Lin Bao$^{1,3,4}$},
  \textbf{Zongwei Wang$^{1,3,5}$},
  \textbf{Yimao Cai$^{1,3,\dagger}$}
\\
\\
  \textsuperscript{1} Peking University, 
  \textsuperscript{2} Nanjing University,\\
  \textsuperscript{3} Beijing Advanced Innovation Center for Integrated Circuits, \\
  \textsuperscript{4} Beijing University of Posts and Telecommunications, 
  \textsuperscript{5} Yanxin Co. Ltd.
}

\begin{document}
\maketitle

\begin{abstract}
Tackling complex reasoning tasks typically relies on massive monolithic LLMs, which suffer from severe computational redundancy. While task decomposition through structured pipelines or multi-agent collaborations offers an alternative, these approaches inevitably fall into a critical dilemma: predefined static topologies are highly vulnerable to cascading errors, whereas unconstrained dynamic agents suffer from trajectory divergence and unpredictable memory bloat. To address this, we present \textit{DynaGraph}, a lightweight multi-model framework driven by dynamic topological reconfiguration. At the execution level, \textit{DynaGraph} multiplexes time-division PEFT adapters over a shared base model, enabling both full system training and inference deployment on a single consumer-grade GPU. At the routing level, the Evaluator continuously monitors execution confidence to trigger hierarchical self-healing: \textit{Fine-grained Patching} for localized data gaps and \textit{Subgraph Reconstruction} for severe logical ruptures.  Experiments on StrategyQA, MATH, and FinQA demonstrate our 8B model closely approximates the reasoning capabilities of a 72B monolithic model (e.g., 87.6\% on StrategyQA, 82.7\% on MATH). Furthermore, it reduces latency by up to 68.1\% and token consumption by 68.6\% compared to unconstrained dynamic architectures.
\end{abstract}

\section{Introduction}
\label{tex:intro}

In recent years, LLMs have demonstrated unprecedented capabilities in the field of natural language processing \cite{dubey2024llama3}. 
However, as technology advances, the application scenarios of LLMs have expanded to complex tasks requiring deep logical deduction, such as multi-hop question answering, mathematical reasoning, and financial risk analysis \cite{yang2018hotpotqa, hendrycks2021measuring, wu2023bloomber}. 
Currently, solving such tasks typically relies on extremely large-scale monolithic models, such as GPT-4 (estimated 1.8T parameters) \cite{openai2023gpt4}, Claude 3.5 Sonnet, or Llama 3-70B \cite{dubey2024llama3}. 
Unfortunately, relying on a single massive model to handle all segments within a task pipeline not only introduces severe computational redundancy but also limits the system's flexibility for modular expansion of specific capabilities.

To mitigate the limitations of monolithic models on complex tasks, existing solutions primarily focus on augmenting the single-model paradigm. Prompting techniques like CoT \cite{wei2022chain} and ToT \cite{yao2023tree} guide intermediate reasoning, while RAG \cite{asai2024selfrag} integrates external knowledge. Alternatively, MoE \cite{jiang2024mixtral} scales capacity via sparse activation. However, these methods face architectural bottlenecks. Despite heuristic pruning, prompt-based paradigms operate within predefined, rigid topologies lacking reactive runtime self-healing, making them vulnerable to cascading failures from localized anomalies. Meanwhile, MoE remains a highly coupled system demanding steep memory overhead, which prevents independent modular hot updates.

Consequently, decomposing tasks into multi-agent systems has emerged as an alternative \cite{hong2024metagpt}. However, current collaborative frameworks struggle to balance controllability and flexibility, typically falling into a dilemma between two extremes: \textbf{1) Static DAG Pipelines:} These offer predictable overheads but lack reactive recovery mechanisms. Localized errors inevitably accumulate along the task chain, leading to severe cascading failures \cite{zhu2025where}. \textbf{2) Unconstrained Dynamic Reflection:} Autoregressive agents exhibit high autonomy but are prone to trajectory divergence and meta-cognitive hallucinations \cite{huang2024large, lu2026auditing}. Their repeated trial-and-error leads to unpredictable computational costs and severe GPU memory bloat. Thus, developing a collaborative system that achieves convergent, efficient self-correction under a strictly controlled budget remains a critical challenge.

To address these challenges, we present \textit{DynaGraph}, a lightweight multi-model system with dynamic topological reconfiguration. The framework shifts from static task scheduling to adaptive topological evolution via a real-time feedback loop. At the execution layer, \textit{DynaGraph} multiplexes LoRA adapters over a shared backbone, maintaining a constant $\mathcal{O}(1)$ GPU memory footprint regardless of expert pool size. At the control layer, the Evaluator monitors real-time execution confidence, instantly halting flawed reasoning steps. Based on error severity, it adaptively triggers hierarchical interventions: Fine-grained Patching to suture localized information gaps, or Subgraph Reconstruction to truncate and regenerate severely corrupted branches. This paradigm endows the system with robust self-healing, ensuring trajectory convergence under strict operational budgets.

In summary, our contributions are three-fold: 
\begin{itemize}
    \setlength{\itemsep}{0pt}      
    \setlength{\parskip}{0pt}      
    \setlength{\parsep}{0pt}       
    \setlength{\topsep}{0pt}       
    \setlength{\partopsep}{0pt}    
    \item \textbf{Lightweight Architecture:} \textit{DynaGraph} multiplexes time-division PEFT adapters over a shared base model. This bounds the peak memory footprint to 16.6\,GB, enabling complex multi-model inference on a single consumer-grade GPU.
    
    \item \textbf{Adaptive Topological Self-Healing:} A state-aware reconfiguration mechanism balances execution structure and autonomy. By dynamically deploying \textit{Fine-grained Patching} to suture localized data gaps and \textit{Subgraph Reconstruction} to truncate and regenerate fatally corrupted logic branches, the system guarantees reasoning convergence while halting cascading errors.
    
    \item \textbf{Efficiency and Performance Breakthrough:} Our 8B-scale \textit{DynaGraph} achieves superior performance (e.g., 87.6\% on StrategyQA, 82.7\% on MATH), closely approximating 72B massive models. Compared to unconstrained dynamic architectures like Reflexion, it reduces token consumption and  latency by up to 68.6\% and 68.1\%.
\end{itemize}
\section{Background}
\label{tex:background}

\subsection{Large Language Models}
Large Language Models (LLMs) (e.g., Gemini~\cite{geminiteam2023gemini} and Qwen~\cite{yang2025qwen}) exhibit remarkable capabilities across domains like mathematical reasoning~\cite{cobbe2021training} and commonsense question answering~\cite{talmor2019commonsenseqa}, propelled by scaling laws~\cite{kaplan2020scaling} and emergent abilities~\cite{wei2022emergent}. However, monolithic LLMs often struggle with complex multi-domain tasks. To address this, models are frequently fine-tuned on specialized datasets. For instance, domain-specific fine-tuning in financial~\cite{wu2023bloomber}, biomedical~\cite{lee2020biobert}, and legal~\cite{chalkidis2020legalbert} fields yields superior specialized performance without sacrificing general capabilities. These successes inspire the use of parameter-efficient fine-tuning (PEFT) methods, such as prompt tuning~\cite{li2025survey} and LoRA~\cite{hu2022lora}, for efficient domain adaptation.

\begin{figure*}[t]
  \centering
  \includegraphics[width=\textwidth]{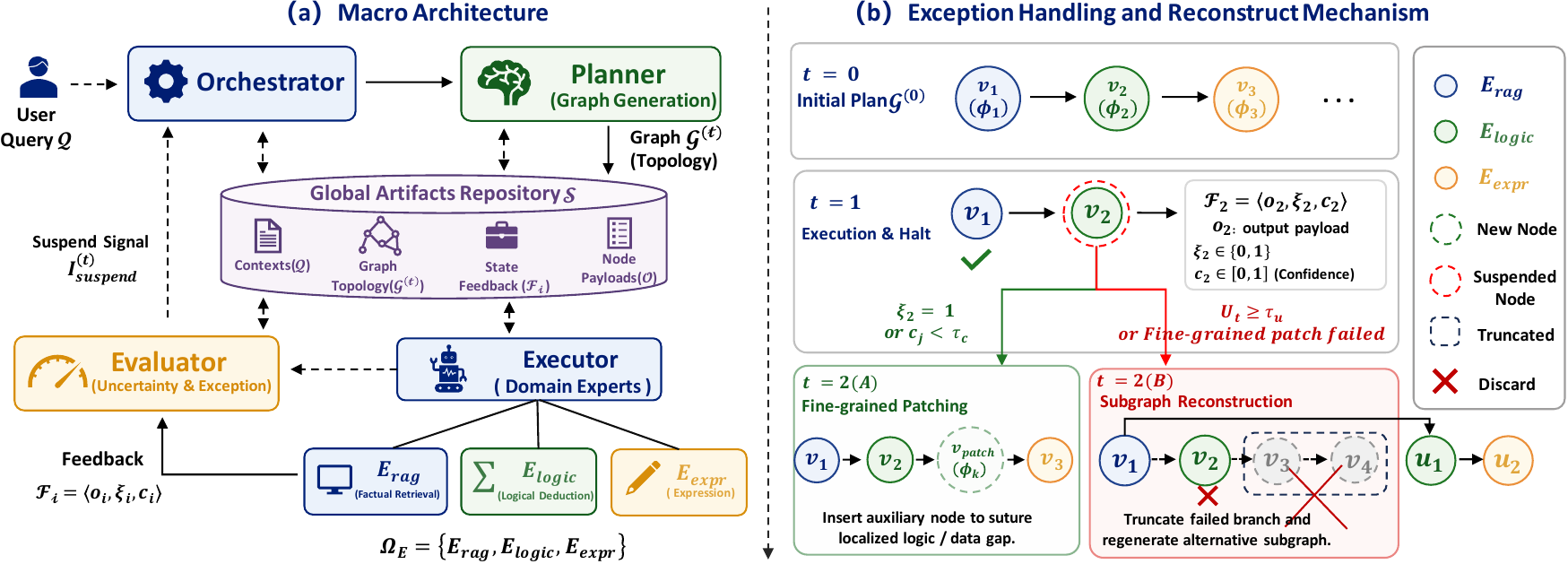}
  \vspace{-6mm}
  \caption{\textbf{Macro architecture and dynamic topology evolution of \textit{DynaGraph}.} 
  (a) \textbf{Macro Architecture} illustrating component interactions.
  (b) \textbf{Exception Handling and Reconstruct Mechanism} through dynamic node insertion and branch replacement.}
  \label{fig:method2}
  \vspace{-4mm}
\end{figure*}

\subsection{Model Interaction}
\subsubsection{Single-Model-Based Interaction}
To enhance reasoning within a single-model paradigm, techniques like Chain-of-Thought (CoT)~\cite{wei2022chain}, zero-shot triggers~\cite{kojima2022large}, and task-breakdown prompts~\cite{zhou2023least} guide intermediate logical deductions. Furthermore, Retrieval-Augmented Generation (RAG)~\cite{lewis2020retrieval} mitigates hallucinations by grounding outputs in external corpora. Nevertheless, single-model approaches remain vulnerable to biased prompts~\cite{turpin2023language}, error accumulation in long inference chains~\cite{lu2026auditing}, and retrieval-generation misalignment~\cite{huang2025survey}, limiting their efficacy on complex tasks.

\subsubsection{Multi-Model-Based Interaction}
Multi-model interactions overcome single-model bottlenecks by decomposing tasks across distinct reasoning modules. Frameworks like Tree of Thoughts~\cite{yao2023tree} and Graph of Thoughts~\cite{besta2024graph} organize reasoning into structured topologies suitable for branching and backtracking. Concurrently, Mixture-of-Experts (MoE) architectures enhance efficiency and accuracy by routing inputs to relevant experts~\cite{zheng2025dynamo}, especially when incorporating fine-grained specialization~\cite{dai2024deepseekmoe} and instruction-tuning~\cite{shen2024mixture}.

Despite their effectiveness, existing multi-model frameworks construct static topologies that cannot adapt to real-time execution states and lack autonomous self-repair mechanisms. \textit{DynaGraph} directly addresses these limitations by introducing dynamic topological reconstruction, pushing multi-model interaction into a resilient, adaptive frontier.
\section{Method}
\label{tex:method}

\begin{figure*}[t]
  \centering
  \includegraphics[width=\textwidth]{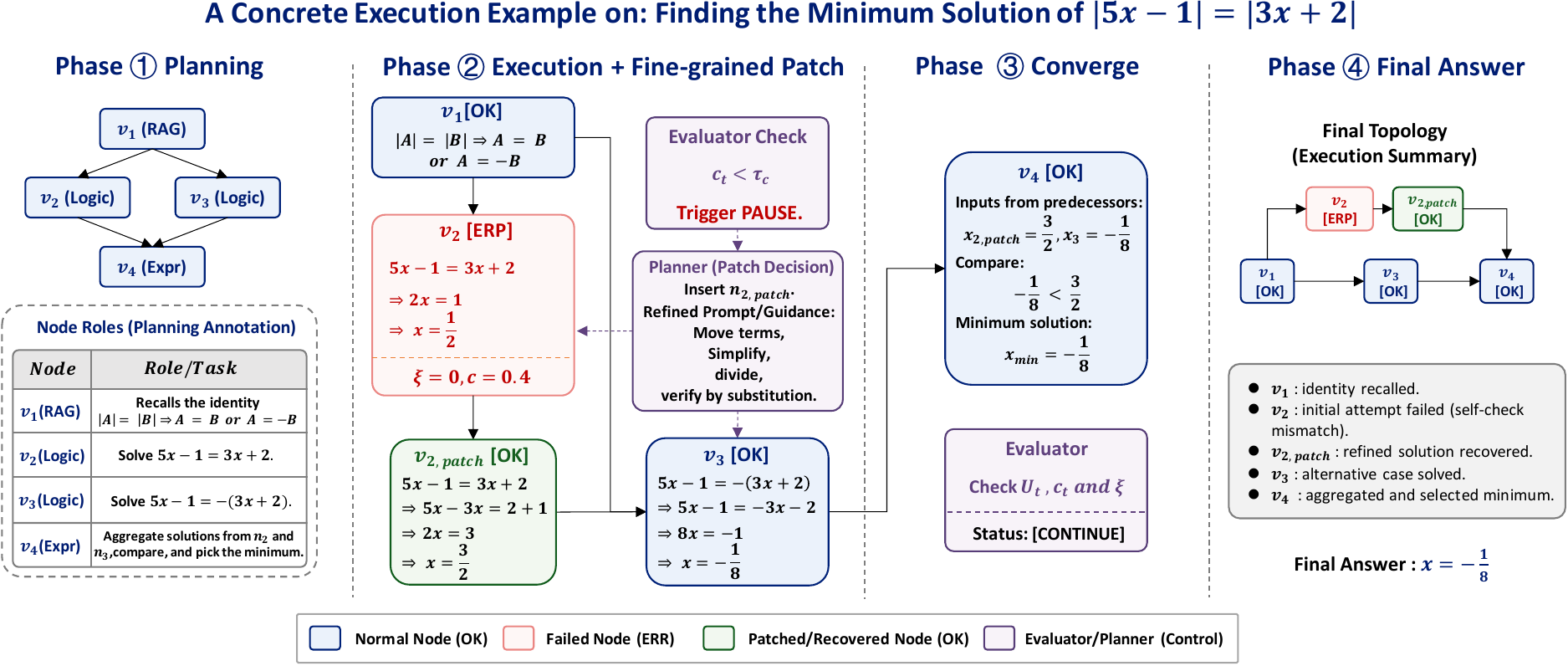}
  \vspace{-6mm}
  \caption{\textbf{A concrete execution example of \textit{DynaGraph}.} The Evaluator halts execution upon detecting anomalies at node $v_2$; a dynamically inserted patch node subsequently recovers the corrupted state to ensure convergence.}
  \label{fig:method1}
  \vspace{-4mm}
\end{figure*}

\subsection{Task Formulation and System Overview}
\subsubsection{Formalization of Task and DAG}

In complex reasoning scenarios, task resolution requires collaborative routing among specialized modules, which we define as the \textit{model interaction topology}.
This paradigm decomposes instructions into a graph where nodes represent expert executions and edges denote contextual dependencies.

We propose this to overcome the inherent limitations of existing frameworks: static pipelines are highly vulnerable to cascading errors, while unconstrained dynamic agents suffer from trajectory divergence.
To balance structured execution with self-healing capabilities, we introduce a \textit{dynamic} topological reconfiguration mechanism.
This allows the system to monitor real-time states and topologically physically restructure the graph to repair logical ruptures and data gaps.

To formalize this dynamic orchestration, given a user query $\mathcal{Q}$, an Orchestrator invokes a Planner to generate an initial task execution graph $\mathcal{G}^{(0)} = (\mathcal{V}^{(0)}, \mathcal{E}^{(0)})$. Unlike static pipelines, our graph topology evolves dynamically with the execution state, defined at time step $t$ as $\mathcal{G}^{(t)} = (\mathcal{V}^{(t)}, \mathcal{E}^{(t)})$.

Each vertex $v_i \in \mathcal{V}^{(t)}$ represents a sub-task node. 
To decouple heterogeneous expert capabilities, we formalize the vertex as Eq.~\eqref{eq:1}, where $E_{\phi_i} \in \Omega_E = \{E_{rag}, E_{logic}, E_{expr}\}$ denotes the assigned domain expert, $q_i$ represents the local instruction context, and $\mathcal{P}_i = \{v_j \mid (v_j, v_i) \in \mathcal{E}^{(t)}\}$ is the set of prerequisite parent nodes. 
Directed edges $e_{j,i} \in \mathcal{E}^{(t)}$ define the logic flow. 
Node $v_i$ is triggered once all prerequisite nodes in $\mathcal{P}_i$ were execute successfully and passed their contexts.
\begin{equation}
v_i = \langle E_{\phi_i}, q_i, \mathcal{P}_i \rangle.
\label{eq:1}
\end{equation}

To systematically record all expert execution outputs, we introduce a structured state space $\mathcal{S}$.
Upon completing its computation, the output state of the assigned expert $E_{\phi_i}$ at node $v_i$ is encapsulated into a standardized feedback tuple, as shown in Eq.~\eqref{eq:2}.
\begin{equation}
\mathcal{F}_i = \langle o_i, \xi_i, c_i \rangle.
\label{eq:2}
\end{equation}
where $o_i$ is the structured output of the corresponding expert. 
The Exception Flag $\xi_i \in \{0, 1\}$ indicates irrecoverable errors (e.g., format deviations or logical ruptures), and $c_i \in [0, 1]$ represents the expert's normalized confidence score. 

\begin{figure*}[t]
  \centering
  \includegraphics[width=\textwidth]{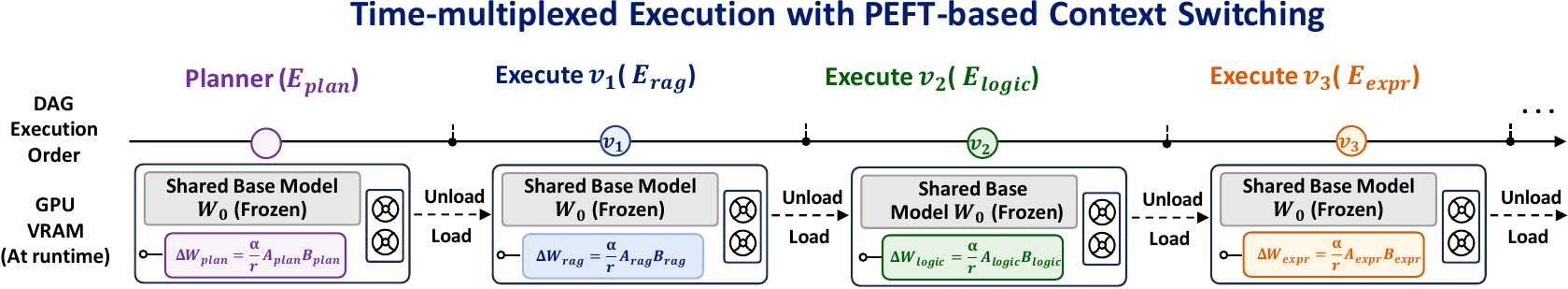}
  \vspace{-6mm}
  \caption{\textbf{Time-multiplexed Execution with PEFT-based Context Switching:} The system maintains a frozen shared base model ($W_0$) and dynamically loads task-specific LoRA adapters (e.g., $A_{rag}B_{rag}$) via PEFT}
  \label{fig:method3}
  \vspace{-3mm}
\end{figure*}

\subsubsection{System Overview}
To address aforementioned problems, we propose \textit{DynaGraph} framework.
As illustrated in Fig.~\ref{fig:method2}~(a), our Framework decouples complex task resolution into four canonical phases: Planning, Execution, Evaluation, and Reconstruction. 
Upon receiving a user query $\mathcal{Q}$, the central Orchestrator prompts the Planner to generate an initial task graph.

Fig.~\ref{fig:method2}~(a) further visualizes the strict module-isolation principle: heterogeneous experts ($E_{\phi_i} \in \Omega_E$) never communicate directly; instead, all intermediate contexts, graph topologies, and node payloads flow through the Global Artifacts Repository $\mathcal{S}$.
As experts return state feedback $\mathcal{F}_i$, the Evaluator monitors the execution stream to quantify uncertainty and detect exceptions. 
If lethal anomalies or high uncertainties exceed thresholds, the Evaluator issues a suspension signal ($\mathbb{I}_{suspend}^{(t)}$), returning control to the Orchestrator. 
As illustrated in Fig.~\ref{fig:method2}~(b), the Planner then executes hierarchical topological reconfiguration: 1) fine-grained patching for localized data errors or gaps (e.g., erroneous intermediate calculations or unretrieved factual entities), or 2) subgraph reconstruction to truncate and regenerate failed branches. 

Fig.~\ref{fig:method1} illustrates a life cycle on finding the minimum solution of $|5x-1|=|3x+2|$. 
\textbf{Phase \ding{172}}: The Planner generates a DAG with nodes $v_1$ through $v_4$. 
\textbf{Phase \ding{173}}: After $v_1$ succeeds, the Evaluator detects a calculation error at $v_2$ and triggers suspension. The Planner inserts a patch node $v_{2,patch}$ to recalculate the faulty nodes. 
\textbf{Phase \ding{174}}: The corrected state from $v_{2,patch}$ propagates downstream alongside the valid sibling output to the aggregation node $v_3$.
\textbf{Phase \ding{175}}: The system emits the final topology and answer, validating that the system can autonomously and dynamically repair faults.

\subsection{Domain-Specific Expert Design}
\subsubsection{Structured State Feedback of Experts}
Given that each expert is tailored for a distinct function, their structured outputs vary accordingly. 
Therefore, we formally define these specific structures below.
This framework constructs an expert pool $\Omega_E = \{E_{rag}, E_{logic}, E_{expr}\}$ characterized by functional orthogonality. 
Upon termination, each node $v_i$ returns a standardized state feedback tuple $\mathcal{F}_i = \langle o_i, \xi_i, c_i \rangle$ to the Orchestrator. 
The specific variables for each expert are constrained as follows:

\textbf{\ding{172} Factual Retrieval Expert $E_{rag}$:} 
Dedicated to open-domain factual verification. 
Its output $o_i = \langle \mathcal{A}, \mathcal{K}, \mathcal{C} \rangle$ comprises the assertion set $\mathcal{A}$, external evidence $\mathcal{K}$, and citation provenance $\mathcal{C}$. 
Confidence $c_i$ measures evidence reliability. 
If retrieval fails or cannot support assertions, $\xi_i = 1$.

\textbf{\ding{173} Logical Deduction Expert $E_{logic}$:} Dedicated to closed-domain logical verification. 
Its output $o_i = \langle \mathcal{H}, \mathcal{V} \rangle$ comprises the reasoning history $\mathcal{H}$ and boolean verification results $\mathcal{V}$. 
Confidence $c_i$ measures logical self-consistency. Upon localized verification failure, $\xi_i = 1$.

\textbf{\ding{174} Expression Expert $E_{expr}$:} Dedicated to linguistic fidelity. 
Its output $o_i = \langle \mathcal{D}, \mathcal{U} \rangle$ contains the generated draft $\mathcal{D}$ and unsupported statements $\mathcal{U}$. 
Confidence $c_i$ measures semantic fidelity. For severe formatting deviations, $\xi_i = 1$.

In practice, the confidence score $c_i \in [0, 1]$ is derived via verbalized self-calibration, prompting experts to explicitly quantify their certainty. The exception flag $\xi_i \in \{0, 1\}$ triggers deterministically if the expert self-reports anomalies or fails structured parsing.

\begin{figure}[t]
  \includegraphics[width=\columnwidth]{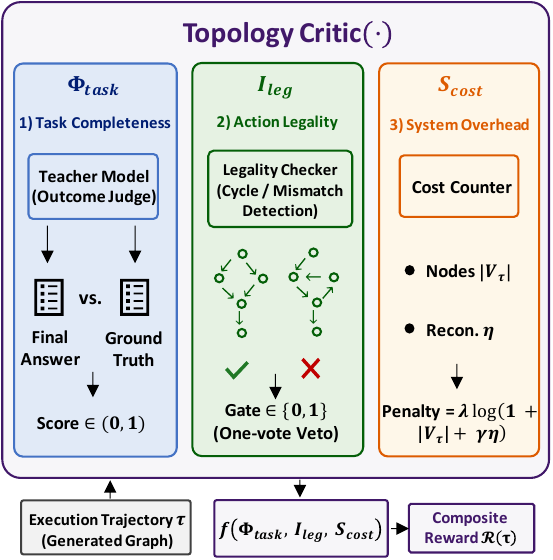}
  \vspace{-5mm}
  \caption{\textbf{Topology Critic:} The execution trajectory $\tau$ is evaluated by a composite reward function $\mathcal{R}(\tau)$ considering task accuracy/completeness, topological legality, and system overhead.}
  \label{fig:method4}
  \vspace{-4mm}
\end{figure}

\begin{figure*}[t]
  \centering
  \includegraphics[width=\textwidth]{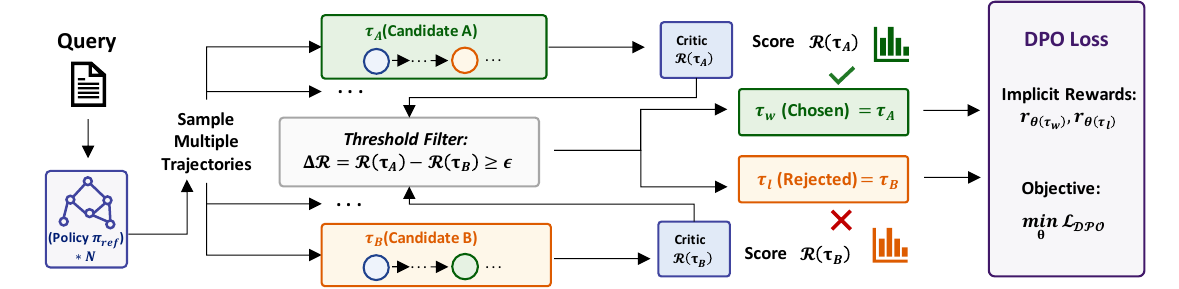}
  \vspace{-6mm}
  \caption{\textbf{Preference Pair Construction \& DPO Optimization:} Candidate trajectories are sampled and ranked by the Critic to construct preference pairs, which are then used to optimize the central Planner via DPO.}
  \label{fig:method5}
  \vspace{-4mm}
\end{figure*}

\subsubsection{Context Switching for Low-Memory Usage}
To address GPU memory bottlenecks, we implement PEFT based time-division scheduling. 
For a frozen base model $W_0 \in \mathbb{R}^{d \times k}$, we train low-rank matrices $A_{\psi} \in \mathbb{R}^{d \times r}$ and $B_{\psi} \in \mathbb{R}^{r \times k}$ ($r \ll \min(d, k)$) for each system module $\psi \in \{E_{plan}, E_{rag}, E_{logic}, E_{expr}\}$.

As illustrated in Fig.~\ref{fig:method3}, during a specific module $\psi$'s execution, the active weights $\theta_{active}$ dynamically incorporate the incremental weights $\Delta W_{\psi}$:
\begin{equation}
\theta_{active} = W_0 + \Delta W_{\psi} = W_0 + \frac{\alpha}{r} (A_{\psi} B_{\psi}).
\label{eq:3}
\end{equation}
where $\alpha$ is a scaling coefficient. 
Fig~\ref{fig:method3} depicts an example of time-division multiplexing timeline: as the DAG execution order advances from the Planner ($\psi=E_{plan}$) to $v_3$ ($\psi=E_{expr}$) and onward, the system atomically unloads the previous LoRA adapter and loads the next. So, at any single time step, the GPU memory contains only the frozen backbone $W_0$ plus one active adapter slice. 

Let $\mathcal{M}(\cdot)$ denote the GPU memory footprint function. The system's peak GPU memory is strictly bounded by Eq.~\eqref{eq:4}.
\begin{equation}
\mathcal{M}_{peak} = \mathcal{M}(W_0) + \max_{\psi} \{\mathcal{M}(A_\psi) + \mathcal{M}(B_\psi)\}.
\label{eq:4}
\end{equation}
Since $r$ is minimal, $\mathcal{M}(A_\psi) + \mathcal{M}(B_\psi)$ is negligible. 
This mathematically guarantees an $\mathcal{O}(1)$ redundant GPU memory complexity regardless of the scaling of both the expert pool and the planning policies.

\subsection{Exception Handling and Reconstruction Mechanism}

To endow the system with topological plasticity, we introduce a continuous monitoring mechanism that detects intermediate anomalies and dynamically restructures the reasoning graph to self-heal.

We define the global uncertainty $\mathcal{U}_t$ across committed nodes $\mathcal{V}_{exec}^{(t)}$ to measure the aggregate execution instability and the cumulative risk of reasoning divergence, as shown in Eq.~\eqref{eq:5}.
\begin{equation}
\mathcal{U}_t = 1 - \frac{1}{|\mathcal{V}_{exec}^{(t)}|} \sum_{v_j \in \mathcal{V}_{exec}^{(t)}} c_j.
\label{eq:5}
\end{equation}
To prevent error propagation, we define the suspension indicator function $\mathbb{I}_{suspend}^{(t)}$ to govern when the system trigger reconstruction. Once $\mathbb{I}_{suspend}^{(t)} = 1$, the Evaluator halts downstream scheduling and initiates the reconfiguration process, shown in Eq.~\eqref{eq:6}.
\begin{equation}
\mathbb{I}_{suspend}^{(t)} = 
\begin{cases} 
1, & \exists v_j \in \mathcal{V}_{exec}^{(t)}, \; \xi_j = 1 \\ 
1, & \exists v_j \in \mathcal{V}_{exec}^{(t)}, \; c_j < \tau_{c} \\ 
1, & \mathcal{U}_t \geq \tau_{u} \\ 
0, & \text{otherwise}
\label{eq:6}
\end{cases}
\end{equation}
where $\tau_u$ defines the global uncertainty tolerance and $\tau_c$ denotes the minimum single-node confidence threshold. 

As illustrated in Fig.~\ref{fig:method2}~(b), the topological transformation operator $\mathcal{T}$ yields $\mathcal{G}^{(t+1)} = \mathcal{T}(\mathcal{G}^{(t)}, \mathcal{F}_{err})$ via two distinct execution paths:

\textbf{\ding{172} Fine-grained Patching Operator}: Repairs localized deficiencies triggered by irrecoverable node errors or confidence floor breaches.  
As depicted in Fig.~\ref{fig:method2}~(b), upon detecting a lethal failure ($\xi_i = 1$) or a single-node confidence floor violation ($c_j < \tau_c$) at intermediate node $v_i$, the Planner dynamically instantiates an auxiliary patch node $v_{i,patch}$ based on the anomalous node's local context, thereby suturing localized breakpoints.

\textbf{\ding{173} Subgraph Reconstruction Operator}: Rectifies macroscopic deviations. 
If a fine-grained patch fails, or the global uncertainty exceeds the tolerance threshold ($\mathcal{U}_t \geq \tau_u$), the system truncates all downstream branches from the failed node and replaces them with a freshly generated subgraph $\mathcal{G}_{sub}$. This discards the corrupted execution state rather than allowing error propagation.

To prevent infinite recursive reconstruction, we impose an operational budget constraint. 
Let $\eta$ be the cumulative reconstruction count and $\Omega_{max}$ the permissible threshold, as shown in Eq.~\eqref{eq:7}.
\begin{equation}
\eta = \sum_{k=1}^{t} \mathbb{I}_{suspend}^{(k)} \leq \Omega_{max}.
\label{eq:7}
\end{equation}
Once $\eta$ reaches $\Omega_{max}$, the system terminates dynamic reconstruction and invokes a fallback mechanism, guaranteeing computational convergence in worst-case scenarios.

\begin{table*}[t!]
\centering
\caption{
\textbf{Main Experimental Results on Heterogeneous Cognitive Tasks.} 
We report task accuracy (Acc, \%), token consumption alongside estimated compute (\#Tok / TFLOPs), and average end-to-end latency (Lat., seconds). \textbf{Bold} indicates the best performance among models in the 8B parameter class, and \underline{underline} denotes the second best (the 72B large model is excluded from this ranking).
}
\scriptsize
\label{tab:main_results}
\setlength{\tabcolsep}{1mm} 
\resizebox{\textwidth}{!}{%
\begin{tabular}{l | ccc | ccc | ccc | c}
\toprule
\toprule
\multirow{2}{*}{\textbf{Method / Architecture}} & \multicolumn{3}{c|}{\textbf{StrategyQA (Open-domain)}} & \multicolumn{3}{c|}{\textbf{MATH (Deduction)}} & \multicolumn{3}{c|}{\textbf{FinQA (Heterogeneous)}} & \textbf{GPU DRAM} \\
\cmidrule{2-10}
~ & \textbf{Acc ($\uparrow$)} & \textbf{\#Tok / TFLOPs ($\downarrow$)} & \textbf{Lat. ($\downarrow$)} & \textbf{Acc ($\uparrow$)} & \textbf{\#Tok / TFLOPs ($\downarrow$)} & \textbf{Lat. ($\downarrow$)} & \textbf{Acc ($\uparrow$)} & \textbf{\#Tok / TFLOPs ($\downarrow$)} & \textbf{Lat. ($\downarrow$)} & \textbf{Usage (GB)} \\
\midrule
\multicolumn{11}{l}{\textit{\textbf{Static Monolithic Baselines (8B Class)}}} \\
\midrule
\rowcolor{orange!20} 
Standard Prompting       & 65.2 & \textbf{210} / \textbf{3.4}    & \textbf{2.5}  & 45.5 & \textbf{350} / \textbf{5.6}    & \textbf{4.2}  & 55.0 & \textbf{420} / \textbf{6.7}    & \textbf{4.8}  & 16.5 \\
\rowcolor{orange!20} 
Standard CoT             & 78.4 & 650 / 10.4     & 8.0   & 65.4 & 1,020 / 16.3   & 11.8  & 70.5 & 1,260 / 20.2   & 14.9  & 16.5 \\
\rowcolor{orange!20} 
Standard ToT             & 85.8 & 3,400 / 54.4   & 39.5  & 80.0 & 5,430 / 86.9   & 67.6  & 80.7 & 6,930 / 110.9  & 76.3  & 16.5 \\
\rowcolor{orange!20} 
Self-Consistency (k=5)   & 83.3 & 1,700 / 27.2   & 19.8  & 78.2 & 2,730 / 43.7   & 34.4  & 76.4 & 3,490 / 55.8   & 38.4  & 16.5 \\
\midrule
\multicolumn{11}{l}{\textit{\textbf{Unconstrained Dynamic Agents (8B Class)}}} \\
\midrule
\rowcolor{yellow!20} 
ReAct                    & 84.5 & 2,480 / 39.7   & 30.5  & 77.6 & 4,310 / 69.0   & 50.0  & 76.2 & 5,080 / 81.3   & 59.5  & 16.5 \\
\rowcolor{yellow!20} 
Reflexion                & \underline{86.2} & 3,890 / 62.2   & 44.5  & \underline{80.5} & 6,160 / 98.6   & 76.4  & \underline{78.8} & 7,690 / 123.0  & 86.9  & 16.5 \\
\rowcolor{yellow!20} 
Multi-Agent (3 $\times$ 8B) & 84.0 & 1,520 / 24.3   & 17.5  & 79.4 & 2,470 / 39.5   & 29.5  & 77.6 & 2,780 / 44.5   & 34.3  & > 49.5 \\
\midrule
\multicolumn{11}{l}{\textit{\textbf{High-Resource Large Model Reference}}} \\
\midrule
\rowcolor{cyan!20}
Qwen-2-72B-Instruct      & \textit{92.5} & \textit{800} / \textit{115.2} & \textit{10.5} & \textit{89.2} & \textit{1,440} / \textit{207.4} & \textit{16.4} & \textit{86.1} & \textit{1,640} / \textit{236.2} & \textit{20.6} & \textit{> 145.0} \\
\midrule
\rowcolor{green!20} 
\textbf{DynaGraph (Ours, 8B)} & \textbf{87.6} & 1,220 / 19.5 & 15.3  & \textbf{82.7} & 2,170 / 34.7  & 24.4  & \textbf{82.5} & 2,480 / 39.7  & 30.2  & 16.6 \textbf{($\mathcal{O}(1)$)} \\
\bottomrule
\bottomrule
\end{tabular}
}
\vspace{-3mm}
\end{table*}

\subsection{Critic Guided Central Model Training}
\subsubsection{Critic Establishment and Evaluation Dimensions}

To train the Planner, we introduce a Topology Critic $\text{Crit}(\cdot)$ to map generated trajectories into learnable reward signals. 
As shown in Fig.~\ref{fig:method4}, $\text{Crit}(\cdot)$ evaluates discrete trajectories $\tau$ using a composite reward function $\mathcal{R}(\tau)$ across three dimensions:

\paragraph{Task Completion $\Phi_{task}$:} Evaluated by a Teacher Model. 
For deterministic tasks, it strictly weights ground-truth matching. 
For open-ended scenarios, it assesses the ratio of valid propositions in draft $\mathcal{D}$ supported by evidence $\mathcal{K}$, heavily penalizing hallucinations $\mathcal{U}$.

\paragraph{Action Legality $\mathbb{I}_{leg}$:} Acts as a strict gating constraint. Cyclic dependencies or mismatched expert assignments forcibly nullify the trajectory score.

\paragraph{System Overhead $S_{cost}$:} Imposes a sub-linear penalty on total node count $|\mathcal{V}_\tau|$ and reconstruction occurrences $\eta$ to suppress redundant paths.

The composite reward is formally defined as Eq.~\eqref{eq:8}, where $\lambda$ scales the overhead penalty and $\gamma$ balances the reconstruction contribution weight.
This equation utilizes a one-vote veto gating mechanism and a logarithmic penalty to permit reasonable expansion while restricting redundancy.
\begin{equation}
\begin{split}
\mathcal{R}(\tau) = &\left( \prod_{v_i \in \tau} \mathbb{I}_{leg}(v_i) \right) \cdot \Phi_{task}(\tau) \\
&- \lambda \cdot \log \big( 1 + |\mathcal{V}_\tau| + \gamma \eta \big).
\label{eq:8}
\end{split}
\end{equation}

\subsubsection{Critic-Guided Direct Preference Optimization}
We employ Direct Preference Optimization (DPO) to fine-tune the low-rank adapter matrices of the central Planner $\pi_{\theta}$, bypassing unstable reward model training.

As illustrated in Fig.~\ref{fig:method5}, given a query $x$, we sample candidate trajectories using a reference policy $\pi_{ref}$ and score them via $\mathcal{R}(\tau)$. 
Fig.~\ref{fig:method5} details the pipeline for constructing preference pairs. For a given query $x$, the reference policy $\pi_{ref}$ generates several candidate trajectories, which the $\text{Crit}(\cdot)$ then scores. For each valid pair, the higher-scoring trajectory is designated as Chosen ($\tau_w$), and the lower-scoring one as Rejected ($\tau_l$). To filter evaluation noise, a sample pair is incorporated into the preference dataset $\mathbb{D} = \{(x^{(i)}, \tau_w^{(i)}, \tau_l^{(i)})\}_{i=1}^N$ only if the reward differential satisfies a margin threshold $\Delta \mathcal{R} \geq \epsilon$.
Based on the Bradley-Terry model, the implicit reward function is defined as Eq.~\eqref{eq:9},
\begin{equation}
r_\theta(\tau|x) = \beta \log \frac{\pi_\theta(\tau|x)}{\pi_{ref}(\tau|x)},
\label{eq:9}
\end{equation}
where $\beta$ controls the KL divergence penalty. 
The final DPO loss minimizes the negative log-likelihood like Eq.~\eqref{eq:10} and $\sigma(\cdot)$ means the logistic function. 
\begin{equation}
\begin{split}
&\mathcal{L}_{DPO}(\pi_\theta; \pi_{ref}) = \\
&\quad -\mathbb{E}_{(x, \tau_w, \tau_l) \sim \mathbb{D}} \Big[ \log \sigma \big( r_\theta(\tau_w|x) - r_\theta(\tau_l|x) \big) \Big].
\label{eq:10}
\end{split}
\end{equation}
\section{Experiments}
\label{tex:experiment}

\subsection{Experimental Setup}
\label{sec:experimental_setup}

We evaluate \textit{DynaGraph} on three datasets: (1) StrategyQA~\cite{geva2021did} for open-domain multi-hop fact retrieval; (2) MATH~\cite{hendrycks2021measuring} for long-horizon logical deduction and topological resilience; and (3) FinQA~\cite{chen2021finqa} for cross-modal multi-expert orchestration. Using DeepSeek-8B as the shared base model with LoRA-parameterized experts, all experiments run on a single RTX 5090 (32GB) GPU to validate the $\mathcal{O}(1)$ spatial complexity.

We compare against representative 8B architectures. Static Monolithic Baselines include Standard Prompting (zero-shot), CoT~\cite{wei2022chain} (intermediate reasoning), Self-Consistency ($k=5$)~\cite{wang2022self} (majority voting over CoT paths), and ToT~\cite{yao2023tree} (tree search over reasoning branches). Unconstrained Dynamic Agents comprise ReAct~\cite{yao2022react} (heuristic trial-and-error), Reflexion~\cite{shinn2023reflexion} (ReAct with textual error-reflection), and Multi-Agent ($3 \times 8B$) (collaboration lacking dynamic memory optimization). Finally, Qwen-2-72B-Instruct~\cite{yang2024qwen2} serves as the 72B High-Resource Reference to evaluate how closely our 8B system approximates massive-parameter capabilities.

\textbf{Hyperparameters \& Metrics.} All 8B models share the same DeepSeek-8B bf16 checkpoint (temperature=0.7, top-p=0.9) on a single RTX 5090 GPU. Experts utilize LoRA (rank=8, $\alpha=16$, dropout=0.05) with thresholds $\tau_c=0.35$ and $\tau_u=0.45$. GPT-5 serves strictly as the offline DPO training teacher. \#Tok sums all prompt and completion tokens (including failed/reconstructed trajectories and planning calls). TFLOPs are calculated as $2 \times \text{parameters} \times \text{tokens}$. Latency reflects single-sample, unbatched wall-clock time, factoring in a $\sim$0.8\,s adapter hot-loading overhead.

\begin{figure*}[t]
  \centering
  \includegraphics[width=\textwidth]{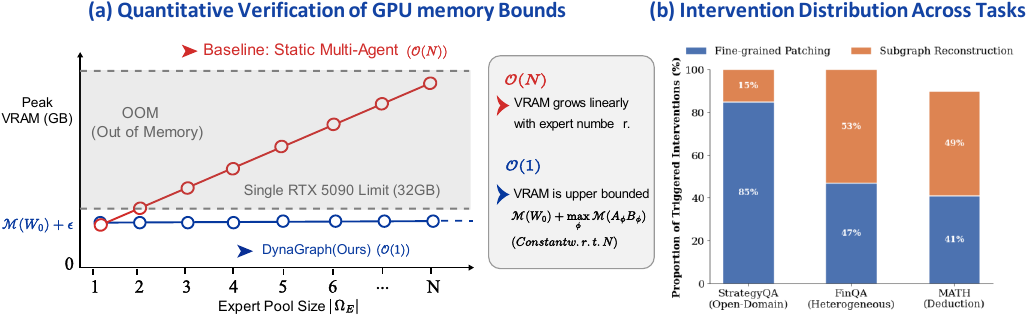}
  \vspace{-6mm}
  \caption{\textbf{System efficiency and adaptive routing of \textit{DynaGraph}.} 
   (a) Quantitative Verification of GPU Memory Bounds: PEFT multiplexing yields a constant $\mathcal{O}(1)$ peak GPU memory footprint. 
    (b) Intervention Distribution Across Tasks: The Controller deploys task-adaptive corrections: StrategyQA predominantly uses \textit{Fine-grained patching} (85\%), while MATH triggers \textit{Subgraph Reconstruction } (49\%) to halt cascading errors.
  }
  \label{fig:system_profiling}
  \vspace{-3mm}
\end{figure*}

\subsection{Main Results: Task Efficacy and Economy}
\label{sec:main_results_analysis}

\subsubsection{Efficacy and High-Resource Approximation}
We evaluate \textit{DynaGraph} across cognitive efficacy and resource constraints, utilizing Tab.~\ref{tab:main_results} and Fig.~\ref{fig:system_profiling}~(a).
As shown in Tab.~\ref{tab:main_results}, \textit{DynaGraph} achieves 87.6\%, 82.7\%, and 82.5\% on StrategyQA, MATH, and FinQA, respectively, outperforming all 8B baselines. Compared to Self-Consistency ($k=5$), it improves average accuracy by approximately 5.0\%. 
Notably, on MATH, the 8B \textit{DynaGraph} (82.7\%) closely approaches the 72B reference (89.2\%), demonstrating that dynamic topological intervention can partially mitigate inherent capability limitations imposed by parameter scale.

\subsubsection{Compute and Time Economy}
Unconstrained dynamic agents and complex tree-search methods suffer severe context redundancy. For instance, Reflexion consumes 7,690 tokens and 86.9\,s on FinQA due to lengthy textual self-reflections, while Standard ToT requires up to 6,930 tokens and 76.3\,s. 
\textit{DynaGraph} replaces linguistic reflection with physical-level graph pruning. While achieving higher accuracy than Reflexion (82.5\% vs. 78.8\% on FinQA), it reduces token consumption by 64.8\%--68.6\% and latency by 65.2\%--68.1\% across benchmarks (e.g., FinQA latency drops from 86.9\,s to 30.2\,s). Furthermore, the TFLOPs metric reveals absolute cross-scale efficiency: despite processing comparable or more tokens than the 72B model, 8B \textit{DynaGraph} incurs only a fraction of its compute .

\subsubsection{Space Economy}
Multi-expert deployment is typically hindered by GPU memory  bottlenecks. The static Multi-Agent ($3 \times 8B$) baseline requires $>49.5$\,GB, posing severe OOM risks. 
In contrast, \textit{DynaGraph} maintains a stable peak GPU memory of $\approx$16.6\,GB (Fig.~\ref{fig:system_profiling}~(a)) regardless of expert pool scale, empirically validating the $\mathcal{O}(1)$ bound derived in Sec.~\ref{tex:method} and confirming consumer-grade deployability.

\subsection{Ablation Study on Adaptive Interventions}
\label{sec:ablation_study}

\subsubsection{Dynamic Routing Distribution}
\label{sec:routing_distribution}
We conduct ablation studies to investigate the dynamic routing behavior and contextual decision-making capacity of the system.
We first evaluate whether anomalies are routed adaptively rather than relying on static heuristics. 
Fig.~\ref{fig:system_profiling}~(b) shows the proportional distribution of fine-grained patching and subgraph reconstruction across tasks.
The results show a clear alignment between intervention strategy and task-specific error tolerance. 
On StrategyQA, 85\% of anomalies are resolved via lightweight ine-grained Patching, leveraging $E_{rag}$ to append missing facts without disrupting execution. 
Conversely, on MATH, 59\% of anomalies trigger Subgraph Reconstruction to prune erroneous branches and halt cascading errors. 
This confirms that \textit{DynaGraph} evaluates error costs rather than relying on hard-coded thresholds.

\subsubsection{Mechanism Ablation}
\label{sec:mechanism_ablation}

\begin{table}[t]
\centering
\caption{\textbf{Ablation Results of Dynamic Intervention Mechanisms.} We report task accuracy (Acc, \%) and average end-to-end latency (Lat., seconds).}
\label{tab:ablation_results}
\scriptsize
\renewcommand{\arraystretch}{1.3}
\setlength{\tabcolsep}{2pt}
\resizebox{\columnwidth}{!}{%
\begin{tabular}{l | cc | cc}
\toprule
\toprule
\multirow{2}{*}{\textbf{Configuration}} & \multicolumn{2}{c|}{\textbf{StrategyQA}} & \multicolumn{2}{c}{\textbf{MATH}} \\
\cmidrule{2-5}
& \textbf{Acc ($\uparrow$)} & \textbf{Lat. ($\downarrow$)} & \textbf{Acc ($\uparrow$)} & \textbf{Lat. ($\downarrow$)} \\
\midrule
\rowcolor{orange!3}
\textbf{w/o Fine-grained Patching} & 86.9 & 29.4 & 81.2 & 35.4 \\
\rowcolor{orange!3}
\textbf{w/o Subgraph Reconstruction} & 78.4 & 13.2 & 41.5 & 16.9 \\
\rowcolor{orange!20}
\textbf{Full \textit{DynaGraph (Ours)}} & \textbf{87.6} & \textbf{15.3} & \textbf{82.7} & \textbf{24.4} \\
\bottomrule
\bottomrule
\end{tabular}
}
\vspace{-4mm}
\end{table}

To quantify each mechanism's contribution, we isolate them in control experiments. Tab.~\ref{tab:ablation_results} confirms their complementary efficiency-precision boundaries. 
Without fine-grained patching, StrategyQA latency escalates from 15.3\,s to 29.4\,s (+92.2\%) as localized gaps force global re-computation, though accuracy remains intact. This highlights patching's pivotal role in execution economy. 
Conversely, removing subgraph reconstruction collapses MATH accuracy from 82.7\% to 41.5\%, as early hallucinations corrupt the execution state beyond local repair. This underscores structural reconstruction's indispensability for topological resilience and deductive coherence.
\section{Conclusion}
We propose \textit{DynaGraph}, a lightweight framework that overcomes the fragility and memory bottlenecks of multi-agent systems. By multiplexing PEFT adapters and employing dynamic topological reconfiguration, it enables complex inference on a single consumer GPU. Empirically, our 8B model rivals 72B monoliths while drastically reducing latency and computational overhead.
\clearpage
\section*{Limitations}
While \textit{DynaGraph} enables efficient self-healing, its time-division multiplexing inherently enforces sequential expert execution. This bounds VRAM to prevent out-of-memory errors but precludes simultaneous execution on parallelizable sub-tasks, introducing switching overhead. Furthermore, topological reconfiguration relies heavily on local expert self-calibration and predefined confidence thresholds. Empirically, evaluations are limited to specific English-centric datasets; generalizing to specialized domains (e.g., biomedical or legal) or morphologically rich languages requires further exploration. Finally, investigating \textit{DynaGraph}'s scaling laws on massive models (e.g., 70B+) to determine if topological interventions yield diminishing returns remains a critical avenue for future work.

\bibliography{ref}
\end{document}